\documentclass{iopart}

\usepackage{hyperref}
\usepackage[sort&compress,longnamesfirst]{natbib}
\usepackage{hypernat}
\usepackage{color}
\usepackage{graphicx}

\begin{document}

\title[IFS wiki]{The integral field spectroscopy (IFS) wiki\footnote[1]{A wiki is a collection of web pages designed to enable anyone with access to contribute or modify content, using a simplified markup language. The collaborative encyclopedia Wikipedia is one of the best-known wikis.}}

\author{M.\ S.\ Westmoquette$^1$ and K.\ M.\ Exter$^2$}
\address{$^1$University College London, Gower Street, London, WC1E 6BT, UK. \mailto{msw@star.ucl.ac.uk}\\
$^2$Instituut voor Sterrenkunde, KULeuven, Celestijnenlaan 200D, 3001 Leuven, Belgium. \mailto{katrinaexter@gmail.com}}
\author{L.\ Christensen$^3$, M.\ Maier$^4$, M.\ Lemoine-Busserolle$^5$, \\ J.\ Turner$^4$, T.\ Marquart$^6$}
\address{
$^3$European Southern Observatory, Alonso de Cordova 3107, Santiago, Chile \\
$^4$Gemini Observatory, Casilla 603, Colina el Pino S/N, La Serena, Chile \\
$^5$Oxford Astrophysics, Denys Wilkinson Building, Keble Road, Oxford,\\ OX1 3RH, UK \\
$^6$Department of Astronomy and Space Physics, Box 515, 75120 Uppsala, Sweden
}

\begin{abstract}
In this article we present the integral field spectroscopy (IFS) wiki site, \htmladdnormallink{http://ifs.wikidot.com}{http://ifs.wikidot.com}; what the wiki is, our motivation for creating it, and a short introduction to IFS. The IFS wiki is designed to be a central repository of information, tips, codes, tools, references, etc., regarding the whole subject of IFS, which is accessible and editable by the whole community. Currently the wiki contains a broad base of information covering topics from current and future integral field spectrographs, to observing, to data reduction and analysis techniques. We encourage everyone who wants to know more about IFS to look at this web-site, and any question you may have you can post from there. And if you have had any experience with IFS yourself, we encourage you to contribute your knowledge and help the site develop its full potential.

Before re-inventing the wheel, consult the wiki...
\end{abstract}


\section{Integral field spectroscopy}
Integral field spectroscopy (IFS; also known as spatially-resolved spectroscopy or 3D spectroscopy) is a technique that allows you to simultaneously gather spectra of the sky over a two-dimensional field-of-view. Regardless of the technique or spectrograph used to obtain the data, the final product is (usually) a data-cube, with axes of $x$, $y$ (or RA, Dec; the two spatial axes) and wavelength (velocity).

IFS attempts to solve the main disadvantages of traditional long-slit spectroscopy. When applied to extended objects (either intrinsically or due to poor seeing), long slit spectroscopy can suffer significantly from: (1) slit losses, either due to mismatches in the slit width and object size or to wavelength-dependent differential atmospheric refraction (DAR); (2) extremely limited position-dependent information (this can be a major problem for complex extended objects such as disturbed galaxies). To solve these problems, an ability to simultaneously record spectra of each part of the extended object is required. Technically, issue (2) can be addressed with a long-slit by stepping the slit across the target and recording separate exposures for each position, but this is plainly very time-inefficient. The need for IFS is therefore clear.

\begin{figure}
\includegraphics[width=1\textwidth]{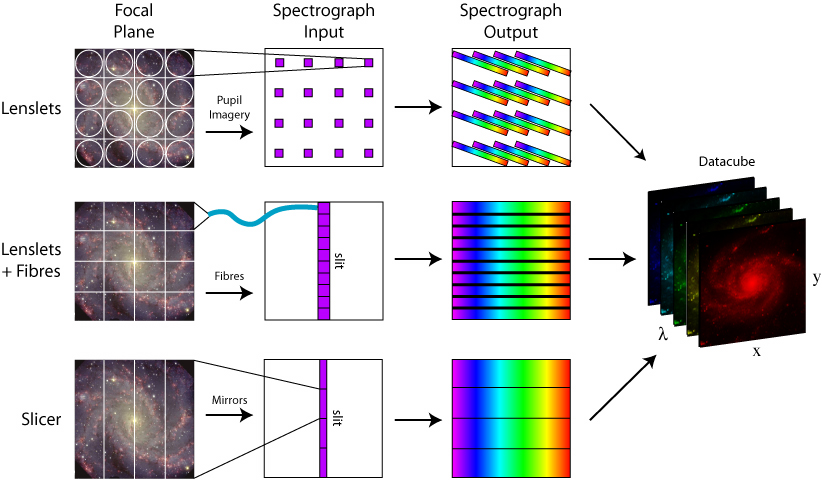}
\caption{The main techniques for achieving integral field spectroscopy. Adapted from \citet{allington98}.}
\label{fig:ifs_techniques}
\end{figure}

An integral field spectrograph consists of two components: the spectrograph and an integral field unit (IFU). The job of the IFU is to divide the 2D spatial plane into an array, and direct the light from each element of the array to the spectrograph. The division of the field-of-view can be achieved in three main ways, as illustrated in Fig.~\ref{fig:ifs_techniques}:

\begin{itemize}
  \item Lenslet array: The input image is split up by a microlens array (MLA). Light from each element of the observed object is then focussed by the microlenses into a small dot and dispersed directly by the spectrograph onto the CCD. Examples: CFHT Tiger, WHT SAURON
  \item Fibres (with or without lenslets): this is currently the most common technique in use. The input image is formed at the front surface of a bundle of optical fibres. The flexibility of the fibres allows the round/rectangular field-of-view to be reformatted into one (or more) `slits', from where the light is directed to the spectrograph and thence on to the CCD. Contiguous lenslets placed in front of the fibres enable the light that would fall in the gaps between the fibres to be collected, resulting in a continuously sampled field-of-view. Examples (without lenslets): WHT Integral; WIYN SparsePak; (with lenslets): VLT FLAMES/ARGUS, Gemini GMOS-IFU
  \item Image-slicer: The input image is formed on a mirror that is segmented in thin horizontal sections, reflecting each 'slice' in slightly different directions. A second segmented mirror is arranged to reformat the slices so that instead of lying on top of each other they lie end to end to form the slit of the spectrograph. This system is especially suitable for the infrared as it uses inherently achromatic mirrors and due to its compactness can easily be contained and cooled to cryogenic temperatures. Examples: VLT SINFONI, Gemini NIFS
\end{itemize}

A less common approach to IFS is the Fabry-P\'{e}rot imaging spectrograph. This allows a large field to be surveyed in a single exposure but only at a single wavelength- the required data volume ($x$, $y$, $\lambda$) is built up by scanning through the desired wavelength range (in a similar fashion to radio receivers), although the wavelength range is necessarily very restricted.

On the most part, the latest IFS instruments are all optimised for use in the optical-red and near-IR. This is due to a number of technical factors and limitations, including the difficulty of manufacturing blue-sensitive optical fibres and the inherently achromaticity of image slicers. Furthermore, since adaptive optics (AO) corrections are more efficient at longer wavelengths, today's large optical telescopes are all becoming red/near-IR optimised.

A fuller introduction to IFS is given in the wiki: \htmladdnormallink{http://ifs.wikidot.com}{http://ifs.wikidot.com}.

\section{Our motivation}
The field of integral field spectroscopy is now well developed, with IFS instruments installed on all the main optical telescope facilities around the world. Excellent work based on IFS data is being published by many groups on subjects varying from planet detection to gravitational lensing. However, IFS continues to be avoided by large sections of the astronomical community due to perceived difficulties with data handling, reduction and analysis. There is no doubt that dealing with IFS data is more complicated than simple imaging or long-slit spectroscopy (indeed whole conferences continue to be dedicated to IFS techniques), but many of the problems that arise could easily be avoided through benefiting from the experience and knowledge of others.

In 2002, the European Commission funded Euro3D Research Training Network (RTN) was set up under the co-ordination of Martin Roth to promote 3D spectroscopy within Europe. Spanning 11 partner institutes, the programme encompassed the promotion of a broad range of research topics, the development of data format standards and a standardised data analysis software package, and the organisation of a number of conferences and training activities. The programme was very successful in promoting IFS within Europe. However, the data format standard has since only been adopted by a limited number of instrument teams (the two large observatories, ESO and Gemini, now follow their own standards), and the analysis package (the E3D vistool\footnote{http://www.aip.de/Euro3D/E3D/}) was not funded to completion and is difficult to install. Continued development and support for generalised packages and standards for IFS, that outside of instrument-specific efforts, ceased when the RTN came to an end in 2005.

In our experience, this has forced many groups to come up with their own independent solutions to data cube manipulation, visualisation and analysis (in a sense `re-inventing the wheel' each time), and because of the effort needed to do this, the tools developed are not automatically made publicly available. The wide variety of IFS instrument types and science uses certainly limits the applicability of generic or generalised software somewhat, but this makes the sharing of information and tools even more important.

We therefore thought it valuable to have a central repository of information, tips, codes, tools, references, etc., regarding the whole subject of IFS which is accessible and editable by the whole community. To this end we have set up a wiki site which we hope will eventually become this information repository: \htmladdnormallink{http://ifs.wikidot.com}{http://ifs.wikidot.com}.

\begin{figure}
\includegraphics[width=1\textwidth]{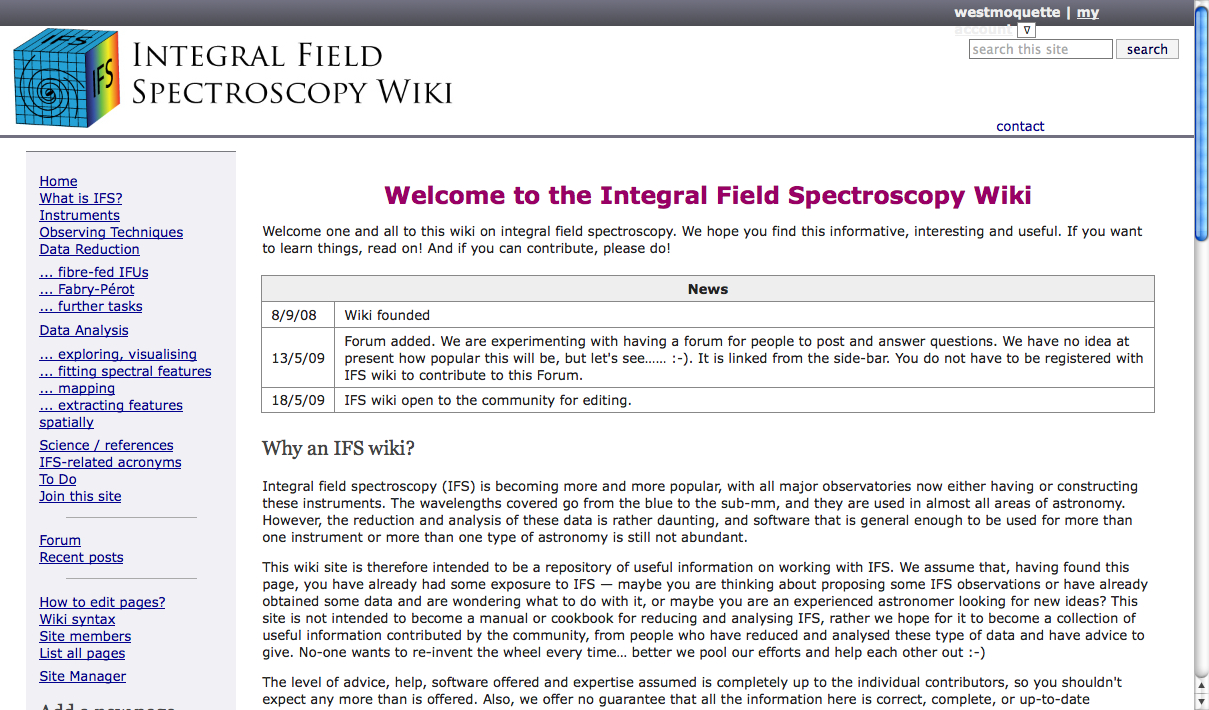}
\caption{A screen-shot of the main welcome page of the wiki}
\label{fig:screen-shot}
\end{figure}

\section{The wiki}
Work on the wiki was begun in 2008. Since then we have sought contributions from experts in the field (the co-authors of this article) in order to build up a solid foundation of content. This first stage ended on the 18th May 2009 when the wiki was opened (made fully editable) to the community. As of now, the wiki covers topics from current and future integral field spectrographs, to observation planning, to data reduction and analysis techniques. A more detailed listing is given below and a screen-shot of the welcome page is shown in Fig.~\ref{fig:screen-shot}.

Building this base of information was essential to make the wiki useful to those with less experience from the word go. If you count yourself in this category and you want to know more about IFS, then please visit the site---make it your first port of call if you have any questions or issues with IFS data you have to work on.

However, if you've had any experience with IFS, please think about contributing your knowledge. For the site to develop its full potential we need the community to make continued contributions: if something is missing or wrong, if you have a useful hint or piece of code that has really helped you with your data, or you have experience of particular instrumental quirks, then please add this to the site. This type of information is of essential use to all the community. Anyone who contributes in any significant way will be listed on the `main contributors' page, and occasional chocolate-based rewards may find their way to supercontributors...

The site will be regularly edited/moderated for structure and English in order to keep it as useable and clear as possible, so contributors who are worried about their English need not worry.

\section{Contents}

As of the time of writing, the wiki contains information under the following headings.

\begin{itemize}
  \item Introduction to integral field spectroscopy
  \item Current and planned (and a few past) instruments
  \item Observing Techniques
  \begin{itemize}
    \item[-] Proposal and Observation Planning
    \item[-] Nod \& Shuffle techniques
    \item[-] AO coupling
    \item[-] Splitting exposures (dithering \& mosaicking)
  \end{itemize}
  \item Data Reduction
  \begin{itemize}
    \item[-] Reducing data from fibre-fed IFUs (aperture tracing, throughput correction, sky/background subtraction, cosmic-ray removal, flux calibration)
    \item[-] Reducing data from image-slicer IFUs
    \item[-] Reducing data from Fabry-P\'{e}rot spectrographs
    \item[-] Advanced reduction tasks (data formats and conversions, mosaicking/combining/resampling/interpolating multiple exposures, differential atmospheric refraction correction, dealing with PSF problems)
    \item[-] Instrument-specific information and data reduction tips
  \end{itemize}
  \item Data Analysis
  \begin{itemize}
    \item[-] Exploring/visualising your data
    \item[-] Spectral fitting
    \item[-] Mapping your results
    \item[-] Extracting sources
  \end{itemize}
  \item A list of science and instrumentation references
  \item Discussion forum
\end{itemize}

At the end of most pages there is space to add comments. This might be a good way to make short contributions, or to post questions or corrections relevant to that specific page. There is also a discussion forum, which would be the place to post more in-depth questions and answers to the community.

Having made this wiki outside of our (paid research) working hours, we would be most grateful for any support you could give to the site---whether that is visiting and making use of the combined knowledge or actually contributing to this knowledge pool.

\bibliographystyle{mn2e}
\bibliography{/Users/msw/Documents/work/references}

\begin{thebibliography}{}

\bibitem[\protect\citeauthoryear{{Allington-Smith}, {Content} \&
  {Haynes}}{{Allington-Smith} et~al.}{1998}]{allington98}
{Allington-Smith} J.~R.,  {Content} R.,    {Haynes} R.,  1998, in {D'Odorico}
  S.,  ed., Proc. SPIE Vol. 3355, p. 196-205, Optical Astronomical
  Instrumentation {New developments in integral field spectroscopy}.
pp 196--205

\end{thebibliography}

\end{document}